\documentclass[aps,prl,superscriptaddress,10pt,onecolumn]{revtex4-2}
\usepackage{graphicx}
\usepackage{amsmath,amsthm,amssymb,dsfont}
\usepackage{mdframed}
\usepackage{orcidlink}
\usepackage{xcolor}

\usepackage{hyperref}

\begin{document}
\title{Uncertainty limits for post-selected metrology}
\author{Ming Ji \orcidlink{0000-0002-6569-5099}}
\email{physmji@gmail.com}
\affiliation{QICI Quantum Information and Computation Initiative, School of Computing and Data Science, The University of Hong Kong, Pokfulam Road, Hong Kong}
\author{Holger F. Hofmann \orcidlink{0000-0001-5649-9718}}
\email{hofmann@hiroshima-u.ac.jp}
\affiliation{Graduate School of Advanced Science and Engineering, Hiroshima University, Kagamiyama 1-3-1, Higashi Hiroshima 739-8530, Japan}

\begin{abstract}
For unitary transformations, the quantum Fisher information (QFI) of a pure state is given by the uncertainty of the generator in that state. In post-selected metrology, the QFI is given by a modified expression describing conditional quantum statistics of the generator. Here, we show that the conditional generator uncertainties defined by post-selected QFI correspond to Ozawa-Hall uncertainties known from the theoretical analysis of quantum measurements. The post-selected measurement outcome updates the generator uncertainty according to the quantum statistics of that outcome. Enhancements of QFI beyond the maximal uncertainties of the generator eigenvalues are possible because post-selection tends to concentrate the largest part of the initial generator uncertainty in low probability outcomes of the post-selection measurement. Anomalous conditional uncertainties thus explain the extreme sensitivities that can be achieved in post-selected metrology. 
\end{abstract}

\maketitle

In quantum metrology, the precision with which the parameter of a unitary transformation can be determined is related to quantum coherence in the eigenstate basis of the unitary ~\cite{Giovannetti2004Quantum,Giovannetti2006QUantum}. The estimation precision is quantified by the quantum Fisher information (QFI)~\cite{Giovannetti2011advances,Braunstein1994Statistical,FUJIWARA1995quantum}, which is upper-bounded by the uncertainty of the generator that appears in the definition of the parameterized unitary operation~\cite{pang2014entanglement,Pang2015improving}. The highest QFI is achieved with pure state inputs that maximize the generator uncertainty~\cite{Giovannetti2011advances}. In multi-photon interferometry, the generator is the photon number difference between the paths, and the maximal uncertainty is achieved by so-called NOON states, superpositions of the two configurations where all photons are in the same path. As this example shows, it can be very difficult to generate non-classical states with extreme generator uncertainties. It may therefore be useful to post-select outcomes that have a particularly high phase sensitivity in order to concentrate the available QFI on a smaller number of measurement outcomes. The enhancement of sensitivity by post-selection has been discussed in the context of weak measurements~\cite{Pang2015improving}, where the observation of anomalous weak values~\cite{Aharonov1988Weak,ipsen2022anomalous,wagner2023simple,masiello2026anomalous,xie2026experimental} can be interpreted as the concentration of a measurement signal within a small fraction of post-selected outcomes. However, the relation between post-selected QFI and weak values is not straightforward. Parameter estimation is based on parameterized change in the probability distribution of more than one outcome. This means that the post-selection procedures must be selected carefully. In the first experimental demonstration of QFI enhancement in post-selected metrology, this was done by applying a post-selection measurement that did not resolve any eigenstates of the system~\cite{Gladstein2022negative}. As shown by Arvindsson-Shukur and coworkers, the enhancement of QFI can be explained in terms of negative quasi-probabilities of an extended Kirkwood-Dirac distribution~\cite{Arvidsson-Shukur2020KD}. These negative probabilities are associated with destructive interferences between eigenstates of the generator in the post-selection probability, a concept that is sufficiently easy to apply in a small Hilbert space~\cite{ji2026enhanced}. Since destructive interference can reduce the post-selection probability to extremely low values while leaving the parameter dependence of the probabilities unchanges, post-selected metrology can achieve a QFI that is much larger than the Heisenberg limit set by the extrema eigenvalues of the generator. This raises two questions. First, it would be good to know whether quantum uncertainties also define a limit for post-selected QFI. Second, it would help to understand what relation between the physics of the state and the physics of the post-selection procedure is responsible for the enhancement. The answers to these two questions should enable a more systematic exploration of post-selected metrology in a wide range of physical systems. 

Regarding the first question, we note that post-selection modifies the statistics of a system by updating the past of the system. Post-selection is theoretically equivalent to a measurement: we choose a specific measurement operator, retain only those experimental runs in which that outcome is obtained, and the state of the resulting subsystem is then defined by a combination of the initial state vector and the measurement operator. In general, one would expect that this additional specification of the system would reduce all uncertainties. However, Bayesian updates for low probability outcomes can increase uncertainties by reducing an initial state bias in the statistics of the generator observable. If we can identify the conditional uncertainties for an arbitrary combination of initial state and post-selection condition, we can expect this uncertainty to accurately describe the upper bound of post-selected QFI. Extreme enhancements of sensitivity can then be achieved whenever anomalous conditional uncertainties are obtained in the post-selection.

It has been noted in~\cite{hofmann2011uncertainty} that, in the absence of post-selection, the QFI for a specific measurement is limited by the reduction of the generator uncertainty described by the measurement theory of Ozawa~\cite{ozawa2003universally}. Here, the weak value for each measurement outcome can be identified with an optimal estimate of the observable that serves a the generator of the unitary~\cite{hall2004prior}. We therefore refer to the conditional uncertainty obtained in this manner as Ozawa-Hall uncertainty. In this paper, we show that the uncertainty limits of post-selected metrology are indeed given by the Ozawa-Hall uncertainties of the post-selected outcomes. By analyzing the decomposition of initial state uncertainties into conditional Ozawa-Hall uncertainties, we then identify the physics that re-distributes uncertainties between different measurement outcomes. It is shown that a Hilbert space description of quantum uncertainties identifies the optimal post-selection strategy for any combination of pure state inputs and generator observables. 

We start our analysis from an initial state of the system $|\psi\rangle$ and a unitary transformation   
\[
\hat{U}(\phi) = \exp(- \frac{i}{\hbar} \hat{A} \phi),
\]
where the operator $\hat{A}$ is a self-adjoint generator and the parameter $\phi$ is an unknown phase shift. The pure state QFI $I_\psi$ for a phase estimate of $\phi$ depends only on the uncertainty $\Delta A^2$ of the generator $\hat{A}$ in the initial state $|\psi \rangle$,
\begin{equation}
    I_\psi = \frac{4}{\hbar^2} \Delta A^2.
\end{equation}
This sensitivity is obtained by optimally measuring the output statistics of a large ensemble. In post-selected metrology, only the fraction of the ensemble corresponding to a measurement outcome $\hat{E}_m$ is used in the phase estimation procedure. Here, $\hat{E}_m$ is part of a positive operator valued measure (POVM) with $\sum_m \hat{E}_m = \hat{1}$. In principle, we can post-select {\it any} outcome $m$ of the POVM. Since the total sensitivity achieved by all outcomes $m$ is bounded by the uncertainty of $\hat{A}$ in the initial state, the post-selected QFI $I_{\mathrm{ps}}(m)$ is bounded by~\cite{Combes2014limits}
\begin{align}
    \label{eq:original}
    I_{\mathrm{ps}}(m) \leq \frac{1}{P(m)}\frac{4}{\hbar^2}\Delta A^2,
\end{align}
where $P(m)=\langle \psi | \hat{E}_{m} | \psi \rangle$ is the post-selection probability of $m$. As known from the literatures \cite{Arvidsson-Shukur2020KD,Arvidsson2024properties}, the post-selected QFI can be given by 
\begin{align} 
\label{eq:enhancedQFI}
I_{\mathrm{ps}}(m)=&\frac{4}{\hbar^2}\left(\frac{\langle \psi | \hat{A} \hat{E}_m \hat{A} | \psi \rangle}{\langle \psi | \hat{E}_m | \psi \rangle} - 
\left|\frac{\langle \psi | \hat{E}_m \hat{A} | \psi \rangle}{\langle \psi | \hat{E}_m | \psi \rangle}\right|^2\right),
\end{align}
The right hand side of this equation does look similar to a quantum expression of conditional uncertainty. It may therefore be possible to establish a relation between the post-selected QFI and the conditional uncertainties associated with the different outcomes $m$. In general, conditional uncertainties $\epsilon_A^2(m)$ are related to the total uncertainty $\Delta A^2$ of a state by a sum of the average conditional uncertainty and the variance of the conditional averages $A(m)$,
\begin{equation}
\label{eq:decompose}
\Delta A^2 = \sum_m \left( \epsilon^2_A(m) + (A(m) - \langle \hat{A} \rangle)^2 \right) P(m),
\end{equation}
where $P(m)$ is the measurement probability of the outcome $m$ given by $\langle \psi | \hat{E}_m | \psi \rangle$. In quantum mechanics, we can identify the conditional uncertainty with the optimal Ozawa-Hall uncertainty~\cite{iinuma2016experimental} 
\begin{equation}
    \epsilon^2_A(m) = \frac{\langle \psi | \hat{A} \hat{E}_m \hat{A} | \psi \rangle}{\langle \psi | \hat{E}_m | \psi \rangle} - \left(\mbox{Re}
\left(\frac{\langle \psi | \hat{E}_m \hat{A} | \psi \rangle}{\langle \psi | \hat{E}_m | \psi \rangle}\right) \right)^2
\end{equation}
and the conditional averages with the real parts of the weak values,
\begin{equation}
    A(m) = \mbox{Re}
\left(\frac{\langle \psi | \hat{E}_m \hat{A} | \psi \rangle}{\langle \psi | \hat{E}_m | \psi \rangle}\right).
\end{equation}
With these definitions, the QFI of post-selected metrology is thus associated with the optimal Ozawa-Hall uncertainty in terms of 
\begin{equation}
\label{eq:conditional}
    I_{\mathrm{ps}}(m) = \frac{4}{\hbar^2}\left(\epsilon_A^2(m) - \left(\mbox{Im}
\left(\frac{\langle \psi | \hat{E}_m \hat{A} | \psi \rangle}{\langle \psi | \hat{E}_m | \psi \rangle}\right) \right)^2\right).
\end{equation}
The imaginary weak value represents the phase information obtained from $p(m)$. We can use Eq.~(\ref{eq:decompose}) to identify the fraction of the original QFI of $\Delta A^2$ that is lost due to post-selection,
\begin{align}
    \label{eq:original}
    \sum_m I_{\mathrm{ps}}(m) P(m) = \frac{4}{\hbar^2}\left(\Delta A^2 - \sum_m \left| \frac{\langle \psi | \hat{E}_m \hat{A} | \psi \rangle}{\langle \psi | \hat{E}_m | \psi \rangle} - \langle \hat{A} \rangle \right|^2P(m)\right).
\end{align}
For optimal post-selection strategies, the complex weak value for each outcome $m$ should be equal to the expectation value of the state. In general, the average post-selected QFI is lower than the initial QFI by
\begin{equation}
I_0 -  \sum_m I_{\mathrm{ps}}(m) P(m) =  \frac{4}{\hbar^2}\sum_m \left| \frac{\langle \psi | \hat{E}_m \hat{A} | \psi \rangle}{\langle \psi | \hat{E}_m | \psi \rangle} - \langle \hat{A} \rangle \right|^2 P(m).
\end{equation}
It should be stressed that this relation makes intuitive sense - the purpose of the post-selection is to re-distribute uncertainty, not to reduce it. If the conditional value of $\hat{A}$ for $m$ is different from the initial expectation value, the observation of $m$ reduces the uncertainty of $\hat{A}$ accordingly. Such updates should be avoided in post-selected metrology. 

As shown in Eq.~(\ref{eq:conditional}), the QFI behaves exactly like the conditional uncertainty. Post-selection of $m$ can enhance the QFI if and only if the conditional uncertainty $\epsilon_A^2(m)$ is larger than the original uncertainty $\Delta A^2$. Enhancements beyond the limit of maximal eigenvalues is possible if the Ozawa-Hall uncertainties of the outcomes $m$ permit it. It is therefore important to understand the manner in which initial states and post-selection define the Ozawa-Hall uncertainties. Of particular interest will be the case of anomalous conditional uncertainties, where the uncertainty of a physical property appears to be greater than the maximal uncertainty of its eigenvalues.

There is a simple way in which anomalous conditional uncertainties can be obtained. Quantum fluctuations can be associated with a state vector $| \Delta \rangle$ orthogonal to the initial state $| \psi \rangle$, so that
\begin{equation}
    \hat{A} | \psi \rangle = \langle \hat{A} \rangle | \psi \rangle + \Delta A | \Delta \rangle.
\end{equation}
The complex weak value is then given by
\begin{equation}
\frac{\langle \psi | \hat{E}_m \hat{A} | \psi \rangle}{\langle \psi | \hat{E}_m | \psi \rangle} = \langle \hat{A} \rangle + \Delta A \frac{\langle \psi | \hat{E}_m | \Delta \rangle}{\langle \psi | \hat{E}_m | \psi \rangle}.
\end{equation}
If we consider only $\hat{E}_m$ with $\langle \psi | \hat{E}_m | \Delta \rangle=0$, the conditional averages are equal to the initial expectation value and no information is gained about the value of $\hat{A}$. The Ozawa-Hall uncertainty is then given by
\begin{equation}
    \epsilon_A^2(m) = \Delta A^2 \frac{\langle \Delta | \hat{E}_m | \Delta \rangle}{\langle \psi | \hat{E}_m | \psi \rangle}.
\end{equation}
Enhanced QFI can be obtained with $\hat{E}_m|\Delta\rangle =|\Delta \rangle$ and $\langle \psi|\hat{E}_m|\psi \rangle \ll 1$. This relation between quantum fluctuations of $\hat{A}$ and enhanced QFI indicates that the orthogonality of the fluctuation state $| \Delta \rangle$ and the actual input state $| \psi \rangle$ permits a post-selection that isolates and compresses the fluctuations into a low probability part of the statistics.

It may be helpful to consider a simple example inspired by the first experimental demonstration of post-selected metrology~\cite{Gladstein2022negative}. In a two level system, the initial state $| \psi \rangle$ and the fluctuation state $| \Delta \rangle$ form a complete orthogonal basis. For $\Delta A^2=1$, the generator is given by 
\begin{equation}
\hat{A}=| \psi \rangle\langle \Delta | + | \Delta \rangle \langle \psi |.
\end{equation}
We can now compress the uncertainty into a single outcome
\begin{equation}
\hat{E}_0 = | \Delta \rangle\langle \Delta | + \eta | \psi \rangle\langle \psi |.
\end{equation}
With this definition, the Ozawa-Hall uncertainty of the outcome $\hat{E}_0$ achieves its maximal possible value, 
\begin{equation}
    \epsilon_A^2(0) = \frac{\Delta A^2}{\eta},
\end{equation}
where $\eta$ is also the post-selection probability. 
The post-selection compresses all of the fluctuations of $\hat{A}$ into a single low probability outcome. The QFI gets compressed in the same manner, resulting in an enhancement by a factor of $1/P(0)=1/\eta$. 

In principle, the mechanism by which post-selection enhances QFI is closely related to the observation of anomalous weak values. However, parameter estimation cannot be performed after post-selection of a pure state component, since this kind of post-selection necessarily erases all information about the parameter carried by the quantum state. It is therefore not immediately obvious whether the observation of anomalous Ozawa-Hall uncertainties is sufficient for a corresponding enhancement of QFI. In Ref.~\cite{Fukuda2026experimental}, it was noted that the observation of an anomalous Ozawa-Hall uncertainty of particle localization could be used to observe enhanced QFI when the low probability output of a two-path interferometer is selected. We can now explain how  the anomalous Ozawa-Hall uncertainty of a single post-selected output can be harnessed using an ancillary qubit system.

Consider a situation where the outcome $| f \rangle$ has a probability of zero for $\phi=0$. The phase dependence of the amplitude of $|f \rangle$ is given by
\begin{equation}
  \left.  \frac{\partial}{\partial \phi} \langle f | \hat{U}(\phi) | \psi \rangle \right|_{\phi=0}= - \frac{i}{\hbar} \langle f | \hat{A} | \psi \rangle. 
\end{equation}
We can now modify the situation by interacting the ancillary qubit state with the system. The qubit state is prepared in an equal superposition of $|+\rangle$ and $|-\rangle$, where $|+\rangle$ shifts the phase $\phi$ by $\delta$ in the positive direction, and $|-\rangle$ shifts the phase $\phi$ by the same amount in the negative direction. The output state is then given by
\begin{equation}
    | \psi_{\mathrm{active}} \rangle = \frac{1}{\sqrt{2}} \left(|+\rangle\otimes\hat{U}(+\delta)|\psi\rangle + |-\rangle\otimes\hat{U}(-\delta)|\psi\rangle\right).
\end{equation}
The weak values of $\hat{A}$ have opposite sign for $|+\rangle$ and for $|-\rangle$, with
\begin{equation}
\label{eq:extremewv}
\frac{\langle +;f |\hat{A}|\psi_{\mathrm{active}}\rangle}{\langle +;f |\psi_{\mathrm{active}}\rangle} = -  \frac{\langle -;f |\hat{A}|\psi_{\mathrm{active}}\rangle}{\langle -;f |\psi_{\mathrm{active}}\rangle}= i \frac{\hbar}{\delta}. 
\end{equation}
The anomalous imaginary weak values near $P(f)=0$ describe the post-selected QFI observed in measurements of the ancillary qubit system. For $\hat{E}_f=\hat{1}\otimes|f\rangle\langle f|$,
\begin{equation}
\mbox{Im}
\left(\frac{\langle \psi_{\mathrm{active}} | \hat{E}_f \hat{A} | \psi_{\mathrm{active}} \rangle}{\langle \psi_{\mathrm{active}} | \hat{E}_f | \psi_{\mathrm{active}} \rangle}\right) = 0
\end{equation}
and
\begin{equation}
    \epsilon_A(f)^2 = \frac{\hbar^2}{\delta^2}.
\end{equation}
The post-selected QFI is then given by $I_\mathrm{ps}=4/\delta^2$, limited only by the error-free implementation of the phase shift $\delta$ in the interaction with the ancillary qubit. Note that this procedure can be applied around any interference minimum with $P(f)$ close to zero, as demonstrated in a recent experiment ~\cite{Fukuda2026noon}. This example thus shows how anomalous Ozawa–Hall uncertainties obtained in the post-selection of a pure state $|f \rangle$ can be used as a quantum resource for enhancing parameter estimation precision.

In conclusion, we have successfully generalized the uncertainty limit of quantum metrology to the post-selected case by demonstrating that the post-selected QFI corresponds to the conditional Ozawa-Hall uncertainty of the post-selected ensemble, minus the squared imaginary weak value representing the phase sensitivity of post-selection. It may be worth noting that this result includes the conventional uncertainty limit for a trivial post-selection of all outcomes ($\hat{E}_0=\hat{I}$). Our analysis shows that the enhancement of QFI in post-selected metrology corresponds to the selection of extreme conditional uncertainties of the generator $\hat{A}$. In Hilbert space, the fluctuations of $\hat{A}$ are represented by a vector orthogonal to the state vector, so that a high conditional uncertainty is obtained when only a small component of the initial state is selected together with most or all of the orthogonal state vector representing the uncertainty of $\hat{A}$. As we have noted, this post-selection strategy is also responsible for anomalous weak values, and these in turn represent and possibly explain paradoxical quantum statistics~\cite{Leifer2005pre,Tollaksen2007pre,matthew2014anomalous,avella2017anomalous,kunjwal2019anomalous,Pan2020interference,cohen2026quantum}. The reason why this link between post-selected metrology and anomalous weak values is not immediately apparent is that the analysis of quantum paradoxes tends to focus on post-selections of pure state projections, leaving no room for a metrological advantage in subsequent measurements. We have shown that a straightforward extension of the Hilbert space by a two-level ancilla can solve this problem, demonstrating the fundamental nature of the link between paradoxical quantum statistics and the advantage of post-selected metrology. In summary, we have shown that the uncertainty limit of post-selected metrology is given by the same conditional uncertainties that were previously introduced to characterize measurement errors, providing a deeper insight into the relation between post-selected metrology and non-classical statistics. This insight into the underlying physics will be extremely helpful in the optimization of post-selected metrology for a wide range of possible implementations.

\section*{Acknowledgment}
M. Ji acknowledges support from the National Science Foundation of China via the Excellent Young Scientists Fund (Hong Kong and Macau) Project 12322516, the National Natural Science Foundation of China (NSFC)/Research Grants Council (RGC) Joint Research Scheme via Project N\_HKU7107/24, and the Hong Kong Research Grant Council (RGC) through the General Research Fund (GRF) grant 17302724. H. F. Hofmann acknowledges support from ERATO, Japan Science and Technology Agency (JPMJER2402).

\bibliographystyle{unsrturl}
\bibliography{ref.bib}

@article{Aharonov1988Weak,
  title = {How the result of a measurement of a component of the spin of a spin-1/2 particle can turn out to be 100},
  author = {Aharonov, Yakir and Albert, David Z. and Vaidman, Lev},
  journal = {Phys. Rev. Lett.},
  volume = {60},
  issue = {14},
  pages = {1351--1354},
  numpages = {0},
  year = {1988},
  month = {4},
  publisher = {American Physical Society},
  doi = {10.1103/PhysRevLett.60.1351},
}

@Article{Arvidsson-Shukur2020KD,
author={Arvidsson-Shukur, David R. M.
and Yunger Halpern, Nicole
and Lepage, Hugo V.
and Lasek, Aleksander A.
and Barnes, Crispin H. W.
and Lloyd, Seth},
title={Quantum advantage in postselected metrology},
journal={Nature Communications},
year={2020},
month={Jul},
day={29},
volume={11},
number={1},
pages={3775},
issn={2041-1723},
doi={10.1038/s41467-020-17559-w}
}

@article{Arvidsson2024properties,
doi = {10.1088/1367-2630/ada05d},
url = {https://doi.org/10.1088/1367-2630/ada05d},
year = {2024},
month = {dec},
publisher = {IOP Publishing},
volume = {26},
number = {12},
pages = {121201},
author = {Arvidsson-Shukur, David R M and Braasch Jr, William F and De Bièvre, Stephan and Dressel, Justin and Jordan, Andrew N and Langrenez, Christopher and Lostaglio, Matteo and Lundeen, Jeff S and Halpern, Nicole Yunger},
title = {Properties and applications of the Kirkwood–Dirac distribution},
journal = {New Journal of Physics}
}

@article{avella2017anomalous,
  title = {Anomalous weak values and the violation of a multiple-measurement Leggett-Garg inequality},
  author = {Avella, Alessio and Piacentini, Fabrizio and Borsarelli, Michelangelo and Barbieri, Marco and Gramegna, Marco and Lussana, Rudi and Villa, Federica and Tosi, Alberto and Degiovanni, Ivo Pietro and Genovese, Marco},
  journal = {Phys. Rev. A},
  volume = {96},
  issue = {5},
  pages = {052123},
  numpages = {5},
  year = {2017},
  month = {Nov},
  publisher = {American Physical Society},
  doi = {10.1103/PhysRevA.96.052123},
  url = {https://link.aps.org/doi/10.1103/PhysRevA.96.052123}
}

@article{Braunstein1994Statistical,
  title = {Statistical distance and the geometry of quantum states},
  author = {Braunstein, Samuel L. and Caves, Carlton M.},
  journal = {Phys. Rev. Lett.},
  volume = {72},
  issue = {22},
  pages = {3439--3443},
  numpages = {0},
  year = {1994},
  month = {May},
  publisher = {American Physical Society},
  doi = {10.1103/PhysRevLett.72.3439},
  url = {https://link.aps.org/doi/10.1103/PhysRevLett.72.3439}
}

@misc{cohen2026quantum,
      title={Quantum nonlocal correlations of anomalous weak values}, 
      author={Ron Cohen and Avshalom C. Elitzur and Eliahu Cohen},
      year={2026},
      eprint={2607.01491},
      archivePrefix={arXiv},
      primaryClass={quant-ph},
      url={https://arxiv.org/abs/2607.01491}, 
}

@article{Combes2014limits,
  title = {Quantum limits on postselected, probabilistic quantum metrology},
  author = {Combes, Joshua and Ferrie, Christopher and Jiang, Zhang and Caves, Carlton M.},
  journal = {Phys. Rev. A},
  volume = {89},
  issue = {5},
  pages = {052117},
  numpages = {10},
  year = {2014},
  month = {May},
  publisher = {American Physical Society},
  doi = {10.1103/PhysRevA.89.052117},
  url = {https://link.aps.org/doi/10.1103/PhysRevA.89.052117}
}

@article{Fukuda2026experimental,
doi = {10.1088/1367-2630/ae51b7},
url = {https://doi.org/10.1088/1367-2630/ae51b7},
year = {2026},
month = {mar},
publisher = {IOP Publishing},
volume = {28},
number = {3},
pages = {034515},
author = {Fukuda, Ryuya and Iinuma, Masataka and Matsumoto, Yuto and Hofmann, Holger F},
title = {Experimental evidence for the physical delocalization of individual photons in an interferometer},
journal = {New Journal of Physics}
}

@article{Fukuda2026noon,
doi = {},
url = {},
year = {},
month = {},
publisher = {},
volume = {},
number = {},
pages = {},
author = {Fukuda, Ryuya and Tanaka, Katsuki and Iinuma, Masataka and Hofmann, Holger F},
title = {},
journal = {In preparation}
}

@article{FUJIWARA1995quantum,
title = {Quantum Fisher metric and estimation for pure state models},
journal = {Physics Letters A},
volume = {201},
number = {2},
pages = {119-124},
year = {1995},
issn = {0375-9601},
doi = {https://doi.org/10.1016/0375-9601(95)00269-9},
url = {https://www.sciencedirect.com/science/article/pii/0375960195002699},
author = {Akio Fujiwara and Hiroshi Nagaoka}
}

@article{
Giovannetti2004Quantum,
author = {Vittorio Giovannetti  and Seth Lloyd  and Lorenzo Maccone },
title = {Quantum-Enhanced Measurements: Beating the Standard Quantum Limit},
journal = {Science},
volume = {306},
number = {5700},
pages = {1330-1336},
year = {2004},
doi = {10.1126/science.1104149},
URL = {https://www.science.org/doi/abs/10.1126/science.1104149},
eprint = {https://www.science.org/doi/pdf/10.1126/science.1104149}}

@article{Giovannetti2006QUantum,
  title = {Quantum Metrology},
  author = {Giovannetti, Vittorio and Lloyd, Seth and Maccone, Lorenzo},
  journal = {Phys. Rev. Lett.},
  volume = {96},
  issue = {1},
  pages = {010401},
  numpages = {4},
  year = {2006},
  month = {Jan},
  publisher = {American Physical Society},
  doi = {10.1103/PhysRevLett.96.010401},
  url = {https://link.aps.org/doi/10.1103/PhysRevLett.96.010401}
}

@article{Giovannetti2011advances,
  title = {Advances in quantum metrology},
  author = {Giovannetti, Vittorio and Lloyd, Seth and Maccone, Lorenzo},
  journal = {Nat. Photon.},
  volume = {5},
  issue = {4},
  pages = {222-229},
  numpages = {8},
  year = {2011},
  month = {Apr},
  doi = {10.1038/nphoton.2011.35},
  url = {https://doi.org/10.1038/nphoton.2011.35}
}

@article{Gladstein2022negative,
  title = {Negative Quasiprobabilities Enhance Phase Estimation in Quantum-Optics Experiment},
  author = {Lupu-Gladstein, Noah and Yilmaz, Y. Batuhan and Arvidsson-Shukur, David R. M. and Brodutch, Aharon and Pang, Arthur O. T. and Steinberg, Aephraim M. and Halpern, Nicole Yunger},
  journal = {Phys. Rev. Lett.},
  volume = {128},
  issue = {22},
  pages = {220504},
  numpages = {8},
  year = {2022},
  month = {Jun},
  publisher = {American Physical Society},
  doi = {10.1103/PhysRevLett.128.220504},
  url = {https://link.aps.org/doi/10.1103/PhysRevLett.128.220504}
}

@article{hall2004prior,
  title = {Prior information: How to circumvent the standard joint-measurement uncertainty relation},
  author = {Hall, Michael J. W.},
  journal = {Phys. Rev. A},
  volume = {69},
  issue = {5},
  pages = {052113},
  numpages = {12},
  year = {2004},
  month = {May},
  publisher = {American Physical Society},
  doi = {10.1103/PhysRevA.69.052113},
  url = {https://link.aps.org/doi/10.1103/PhysRevA.69.052113}
}

@article{hofmann2011uncertainty,
  title = {Uncertainty limits for quantum metrology obtained from the statistics of weak measurements},
  author = {Hofmann, Holger F.},
  journal = {Phys. Rev. A},
  volume = {83},
  issue = {2},
  pages = {022106},
  numpages = {5},
  year = {2011},
  month = {Feb},
  publisher = {American Physical Society},
  doi = {10.1103/PhysRevA.83.022106},
  url = {https://link.aps.org/doi/10.1103/PhysRevA.83.022106}
}

@article{iinuma2016experimental,
  title = {Experimental evaluation of nonclassical correlations between measurement outcomes and target observable in a quantum measurement},
  author = {Iinuma, Masataka and Suzuki, Yutaro and Nii, Taiki and Kinoshita, Ryuji and Hofmann, Holger F.},
  journal = {Phys. Rev. A},
  volume = {93},
  issue = {3},
  pages = {032104},
  numpages = {9},
  year = {2016},
  month = {Mar},
  publisher = {American Physical Society},
  doi = {10.1103/PhysRevA.93.032104},
  url = {https://link.aps.org/doi/10.1103/PhysRevA.93.032104}
}

@article{ipsen2022anomalous,
  title = {Anomalous Weak Values are Caused by Disturbance},
  author = {Ipsen, A.C.},
  journal = {Found. Phys.},
  volume = {52},
  issue = {},
  pages = {20},
  numpages = {},
  year = {2022},
  month = {Jan},
  publisher = {Spring Nature},
  doi = {10.1007/s10701-021-00534-w},
  url = {https://doi.org/10.1007/s10701-021-00534-w}
}

@misc{ji2026enhanced,
      title={Enhanced quantum parameter estimation based on the Hardy paradox}, 
      author={Ming Ji and Yuxiang Yang and Holger F. Hofmann},
      year={2026},
      eprint={2601.20602},
      archivePrefix={arXiv},
      primaryClass={quant-ph},
      url={https://arxiv.org/abs/2601.20602}, 
}

@article{kunjwal2019anomalous,
  title = {Anomalous weak values and contextuality: Robustness, tightness, and imaginary parts},
  author = {Kunjwal, Ravi and Lostaglio, Matteo and Pusey, Matthew F.},
  journal = {Phys. Rev. A},
  volume = {100},
  issue = {4},
  pages = {042116},
  numpages = {19},
  year = {2019},
  month = {Oct},
  publisher = {American Physical Society},
  doi = {10.1103/PhysRevA.100.042116},
  url = {https://link.aps.org/doi/10.1103/PhysRevA.100.042116}
}

@article{Leifer2005pre,
  title = {Pre- and Post-Selection Paradoxes and Contextuality in Quantum Mechanics},
  author = {Leifer, M. S. and Spekkens, Robert W.},
  journal = {Phys. Rev. Lett.},
  volume = {95},
  issue = {20},
  pages = {200405},
  numpages = {4},
  year = {2005},
  month = {Nov},
  publisher = {American Physical Society},
  doi = {10.1103/PhysRevLett.95.200405},
}

@article{matthew2014anomalous,
  title = {Anomalous Weak Values Are Proofs of Contextuality},
  author = {Pusey, Matthew F.},
  journal = {Phys. Rev. Lett.},
  volume = {113},
  issue = {20},
  pages = {200401},
  numpages = {5},
  year = {2014},
  month = {Nov},
  publisher = {American Physical Society},
  doi = {10.1103/PhysRevLett.113.200401},
  url = {https://link.aps.org/doi/10.1103/PhysRevLett.113.200401}
}

@misc{masiello2026anomalous,
      title={Anomalous weak values in a generalized Mach-Zehnder interferometer extracted directly from intensity measurements}, 
      author={Ismaele V. Masiello and Hartmut Lemmel and Andreas Dvorak and Stephan Sponar and Yuji Hasegawa},
      year={2026},
      eprint={2606.24798},
      archivePrefix={arXiv},
      primaryClass={quant-ph},
      url={https://arxiv.org/abs/2606.24798}, 
}

@article{ozawa2003universally,
  title = {Universally valid reformulation of the Heisenberg uncertainty principle on noise and disturbance in measurement},
  author = {Ozawa, Masanao},
  journal = {Phys. Rev. A},
  volume = {67},
  issue = {4},
  pages = {042105},
  numpages = {6},
  year = {2003},
  month = {Apr},
  publisher = {American Physical Society},
  doi = {10.1103/PhysRevA.67.042105},
  url = {https://link.aps.org/doi/10.1103/PhysRevA.67.042105}
}

@article{Pan2020interference,
  title = {Interference experiment, anomalous weak value, and Leggett-Garg test of macrorealism},
  author = {Pan, A. K.},
  journal = {Phys. Rev. A},
  volume = {102},
  issue = {3},
  pages = {032206},
  numpages = {7},
  year = {2020},
  month = {Sep},
  publisher = {American Physical Society},
  doi = {10.1103/PhysRevA.102.032206},
  url = {https://link.aps.org/doi/10.1103/PhysRevA.102.032206}
}

@article{pang2014entanglement,
  title = {Entanglement-Assisted Weak Value Amplification},
  author = {Pang, Shengshi and Dressel, Justin and Brun, Todd A.},
  journal = {Phys. Rev. Lett.},
  volume = {113},
  issue = {3},
  pages = {030401},
  numpages = {6},
  year = {2014},
  month = {Jul},
  publisher = {American Physical Society},
  doi = {10.1103/PhysRevLett.113.030401},
  url = {https://link.aps.org/doi/10.1103/PhysRevLett.113.030401}
}

@article{Pang2015improving,
  title = {Improving the Precision of Weak Measurements by Postselection Measurement},
  author = {Pang, Shengshi and Brun, Todd A.},
  journal = {Phys. Rev. Lett.},
  volume = {115},
  issue = {12},
  pages = {120401},
  numpages = {6},
  year = {2015},
  month = {Sep},
  publisher = {American Physical Society},
  doi = {10.1103/PhysRevLett.115.120401},
  url = {https://link.aps.org/doi/10.1103/PhysRevLett.115.120401}
}

@article{Tollaksen2007pre,
doi = {10.1088/1751-8113/40/30/025},
url = {https://doi.org/10.1088/1751-8113/40/30/025},
year = {2007},
month = {jul},
publisher = {},
volume = {40},
number = {30},
pages = {9033},
author = {Tollaksen, Jeff},
title = {Pre- and post-selection, weak values and contextuality},
journal = {Journal of Physics A: Mathematical and Theoretical}
}

@article{wagner2023simple,
  title = {Simple proof that anomalous weak values require coherence},
  author = {Wagner, Rafael and Galv\~ao, Ernesto F.},
  journal = {Phys. Rev. A},
  volume = {108},
  issue = {4},
  pages = {L040202},
  numpages = {6},
  year = {2023},
  month = {Oct},
  publisher = {American Physical Society},
  doi = {10.1103/PhysRevA.108.L040202},
  url = {https://link.aps.org/doi/10.1103/PhysRevA.108.L040202}
}

@misc{xie2026experimental,
      title={Experimental Observation of Anomalous Complementary Weak Values from Correlated Pairwise Two-State Vectors}, 
      author={Qian Xie and Liang Xu and Lijian Zhang},
      year={2026},
      eprint={2607.11284},
      archivePrefix={arXiv},
      primaryClass={quant-ph},
      url={https://arxiv.org/abs/2607.11284}, 
}

\end{document}